\documentclass[sigconf]{acmart}
\usepackage{booktabs}
\usepackage{bm}
\usepackage{amsmath}

\usepackage{amssymb,amsfonts}
\usepackage[]{algpseudocode}                               
\usepackage{algorithm}
\algrenewcommand\textproc{\texttt}
\usepackage{caption}
\usepackage{bigstrut}
\usepackage{subfig}
\usepackage{enumitem}
\usepackage{float}
\usepackage{balance}
\usepackage{multirow}
\usepackage{ulem}
\usepackage{tabularx}

\usepackage{pdfpages}
\usepackage{graphicx} 

\theoremstyle{definition}

\usepackage{threeparttable}
\usepackage{makecell}
\newcommand{\fakeparagraph}[1]{\vspace{1mm}\noindent\textbf{#1.}}

\usepackage{cleveref}  

\graphicspath{{./image/}{../}}

\AtBeginDocument{%
  \providecommand\BibTeX{{%
    \normalfont B\kern-0.5em{\scshape i\kern-0.25em b}\kern-0.8em\TeX}}}

\setcopyright{acmlicensed}
\copyrightyear{2024}
\acmYear{2024}
\acmDOI{XXXXXXX.XXXXXXX}

\acmConference[KDD'24]{}{August 25--29,
  2024}{Barcelona, Spain}

\acmISBN{978-1-4503-XXXX-X/18/06}

\begin{document}

\title{HDLdebugger: Streamlining HDL debugging with Large Language Models}

\author{Xufeng Yao$^*$}
\affiliation{%
\institution{CUHK \& Huawei}
\city{Hong Kong SAR, China}
\country{}
}
\email{xfyao@cse.cuhk.edu.hk}

\author{Haoyang Li$^*$}
\affiliation{%
\institution{Huawei}
\city{Hong Kong SAR, China}
\country{}
}
\email{li.haoyang1@huawei.com}

\author{Tsz Ho Chan$^*$}
\affiliation{%
\institution{Huawei}
\city{Hong Kong SAR, China}
\country{}
}
\email{chantszho1@huawei.com
}

\author{Wenyi Xiao}
\affiliation{%
\institution{Huawei}
\city{Hong Kong SAR, China}
\country{}
}
\email{wxiaoae@cse.ust.hk}

\author{Mingxuan Yuan}
\affiliation{%
\institution{Huawei}
\city{Hong Kong SAR, China}
\country{}
}
\email{Yuan.Mingxuan@huawei.com}

\author{Yu Huang}
\affiliation{%
\institution{HiSilicon}
\city{ShenZhen, China}
\country{}
}
\email{huangyu61@hisilicon.com}

\author{Lei Chen$^\dagger$}
\affiliation{%
\institution{HKUST \& HKSUT (GZ)}
\city{Hong Kong SAR, China}
\country{}
}
\email{leichen@cse.ust.hk}

\author{Bei Yu$^\dagger$}
\affiliation{%
\institution{CUHK}
\city{Hong Kong SAR, China}
\country{}
}
\email{byu@cse.cuhk.edu.hk}

\thanks{$^*$ Equal contribution, $^\dagger$  Corresponding author, \\ This work was completed during Xufeng Yao$'$s internship at Huawei}

\renewcommand{\shortauthors}{Trovato and Tobin, et al.}

\begin{abstract}
In the domain of chip design, Hardware Description Languages (HDLs) play a pivotal role. However, due to the complex syntax of HDLs and the limited availability of online resources, debugging HDL codes remains a difficult and time-intensive task, even for seasoned engineers.
Consequently, there is a pressing need to
develop automated HDL code debugging models,
which can alleviate the burden on hardware engineers.
Despite the strong capabilities of Large Language Models (LLMs) in generating, completing, and debugging software code, their utilization in the specialized field of HDL debugging has been limited and, to date, has not yielded satisfactory results.
In this paper, we propose an LLM-assisted HDL debugging
framework, namely HDLdebugger, which consists of HDL debugging data generation via a reverse engineering approach,
a search engine for retrieval-augmented generation,
and a retrieval-augmented LLM fine-tuning approach.
Through the integration of these components, HDLdebugger can automate and streamline HDL debugging for chip design.
Our comprehensive experiments, conducted on an HDL code dataset sourced from Huawei, reveal that HDLdebugger outperforms 13 cutting-edge LLM baselines, displaying exceptional effectiveness in HDL code debugging.
\end{abstract}





\maketitle
\section{introduction}\label{sec:intro}
Hardware Description Languages (HDLs) are crucial in the realm of chip design, serving as the cornerstone for creating, testing, and implementing digital systems~\cite{gordon1995semantic,cong2011high,zhang2015optimizing}. Due to their critical role, the domain of HDL debugging has received comparatively scant attention. 
Traditional debugging approaches primarily involve manual code correction based on syntactic guidelines, followed by iterative testing through compilers. 
This process, while straightforward for languages such as Python and Java, becomes markedly more complex for HDLs due to their sophisticated syntax and the scarcity of accessible resources online. 
Furthermore, compiler-based testing of HDL code, especially in the context of chip development, is exceptionally time-consuming and resource-intensive.

Despite the high demand in the industry for effective HDL debugging techniques and the promising directions they offer, existing methodologies often fall short in addressing the complexities of the problem. 
For example, the template-based method~\cite{jiang2018shaping,liu2019tbar,monperrus2018automatic,huang2023survey}, a traditional strategy in code debugging, utilizes expert-defined code patterns or heuristics to identify and correct errors. However, this approach is inherently limited, capable of rectifying only those errors with predefined patterns. Consequently, it lacks the flexibility and adaptability necessary to tackle a diverse array of bugs. 

\begin{table}[t]
    \small
    \caption{Pilot Debugging Experiments}
    \begin{tabular}{c|c|c|c}
    \toprule
                     Method & GPT4~\cite{achiam2023gpt} & RTLFixer~\cite{tsai2023rtlfixer} &  VeriGen~\cite{thakur2023verigen} \\ \midrule

    Pass-rate$@$1         &     6.35\%           &  28.35\%   &     1.34\%
    \\
    \bottomrule
    \end{tabular}
    \label{tab:pilot-exp}
\end{table}

Recently, researchers have delved into the direct application of large language models (LLMs) to rectify buggy code.
The underlying hypothesis is that LLMs, pre-trained on extensive repositories of open-source code snippets and text, such as Python, can effectively discern bug patterns and automatically repair buggy code.
To assess the efficacy of current LLM-based approaches in addressing industry-level HDL debugging challenges, we conduct a pilot experiment on three typical methods as shown in~\Cref{tab:pilot-exp}.
Among the methods evaluated, GPT4~\cite{achiam2023gpt} is the current state-of-art LLM, RTLFixer~\cite{tsai2023rtlfixer} leverages retrieval augmented generation (RAG)~\cite{gao2023retrieval} and advanced prompt engineering~\cite{yao2022react}, specifically tailored for HDL debugging tasks.
VeriGen~\cite{thakur2023verigen} is a hardware large language model trained by a self-contained hardware dataset.
Despite these advanced approaches, our observations indicate that none of the methods delivered results that met our criteria for satisfaction in the context of industry-level HDL debugging scenarios.
A primary contributing factor to this shortfall is the insufficiency of HDL code resources for training. 
Consequently, these pre-trained LLMs struggle to accurately comprehend the syntax and functionality inherent to HDL codes.

To tackle the problem, we propose an HDL debugging framework, namely HDLdebugger, which consists of three components, i.e., data generation, search engine, and retrieval-augmented LLM fine-tuning. 
Firstly, the data generation procedure targets overcoming the obstacle of the limited availability of HDL bugs.
Specifically, we employ reverse engineering to insert specific modifications into the original error-free code.
Therefore, we can produce corresponding buggy versions and error messages via compilers, which are used to construct a code database and further fine-tune LLMs. 
Secondly, we propose an effective and efficient search engine, which is supported by the code database constructed by the data generation approach and the document database with various internal HDL documents which contain relevant information for buggy codes.
Given a buggy code and its error message, the search engine retrieves relevant information (i.e., document RAG) and buggy codes (i.e., code RAG) with similar patterns from the document database and code database, respectively.
The document RAG and code RAG are crucial for both the fine-tuning and inference stages of our retrieval-augmented LLM, enhancing the ability of LLMs to comprehensively understand the HDL buggy code and repair it effectively.
Thirdly, to enhance the ability of LLMs to generate accurate code solutions, we propose a novel fine-tuning approach for LLMs. This approach incorporates a self-guided thought generation mechanism and a retrieval-augmented fine-tuning process, significantly improving the LLM's performance in debugging HDL code.

Our contributions are summarized as follows:
\begin{itemize}[leftmargin=12pt]
   \item We introduce an advanced LLM-based HDL debugging framework supporting chip designs in the industry, namely HDLdebugger, which consists of buggy data generation, search engine, and retrieval-augmented LLM fine-tuning. 

   \item To address the scarcity of high-quality HDL debugging training data,   we propose a data generation approach based on reverse engineeering to comprehensively generate diverse and realistic HDL buggy codes with the correct version.

   \item We propose a search engine to create code RAG (resp. doc RAG) for
   HDL buggy code (resp. relevant information) effectively and efficiently, enhancing the fine-tuning and inference of LLMs.

   \item We present a novel retrieval-augmented fine-tuning approach for HDL debugging, which integrates self-guided thought generation with RAG-based fine-tuning strategies.

   \item Extensive experiments on the HDL code dataset from Huawei demonstrate superior performance against 13 state-of-the-art baselines, including GPT4 and various HDL debugging LLMs.

\end{itemize}

\section{Methodology}\label{sec:method}
This section provides a comprehensive overview of the proposed HDLdebugger. 
Initially, we delve into the buggy code generation pipeline, as detailed in Section~\ref{ssec:data-aug}. 
Subsequently, the search engine mechanism tailored for Retrieval-Augmented Generation (RAG) is presented in Section~\ref{ssec:search_engine}. 
The Retrieval-Augmented LLM fine-tuning is elaborated upon in Section~\ref{ssec:LLM-train}. The important notations in our paper are shown in Table \ref{tab:notation}.

\begin{table}[t]	
\centering
\small
\caption{\color{black}Important notations.}
\label{tab:notation}
\begin{tabular}{c|l}
    \toprule
    \textbf{Notation} & \textbf{Description} \\ \midrule
    $b_i$      & Buggy code  \\ 
    $e_j$      & Error id \\ 
    $m_i$      & Error messages for $b_i$  \\
    $c_i$      & Correct code  for $b_i$ \\
    $(d_j, r_j, s_j)$ & Descriptions $d_j$,  reasons $r_j$, and  potential solutions  $s_j$ for $e_j$\\ 
    $D_e, D_c$ & error database, code database \\ 
    $\mathbf{z}^w_i$ &  Keyword vector for buggy code $b_i$ with error message $m_i$ \\ 
    $sim(I_i,I_j)$ & Similarity between buggy codes $I_i$ and $I_j$ \\ 
    $q=(b,e)$ & Code query consisting of buggy code $b$ and error message $m$\\
    $rag^d_i$ & The document RAG based on error message $m_i$ and $b_i$ \\ 
    $rag^c_i$ & The code RAG based on buggy code $b_i$ \\ 
    $p_t$ & Prompt of thought generation \\ 
    $p_c$ & Prompt of buggy code correction \\ 
    $t_i$ & The thought for solving buggy code $b_i$ \\
\bottomrule
\end{tabular}
\end{table}

\begin{figure}[t]
    \centering
    \includegraphics[height=4.4cm]{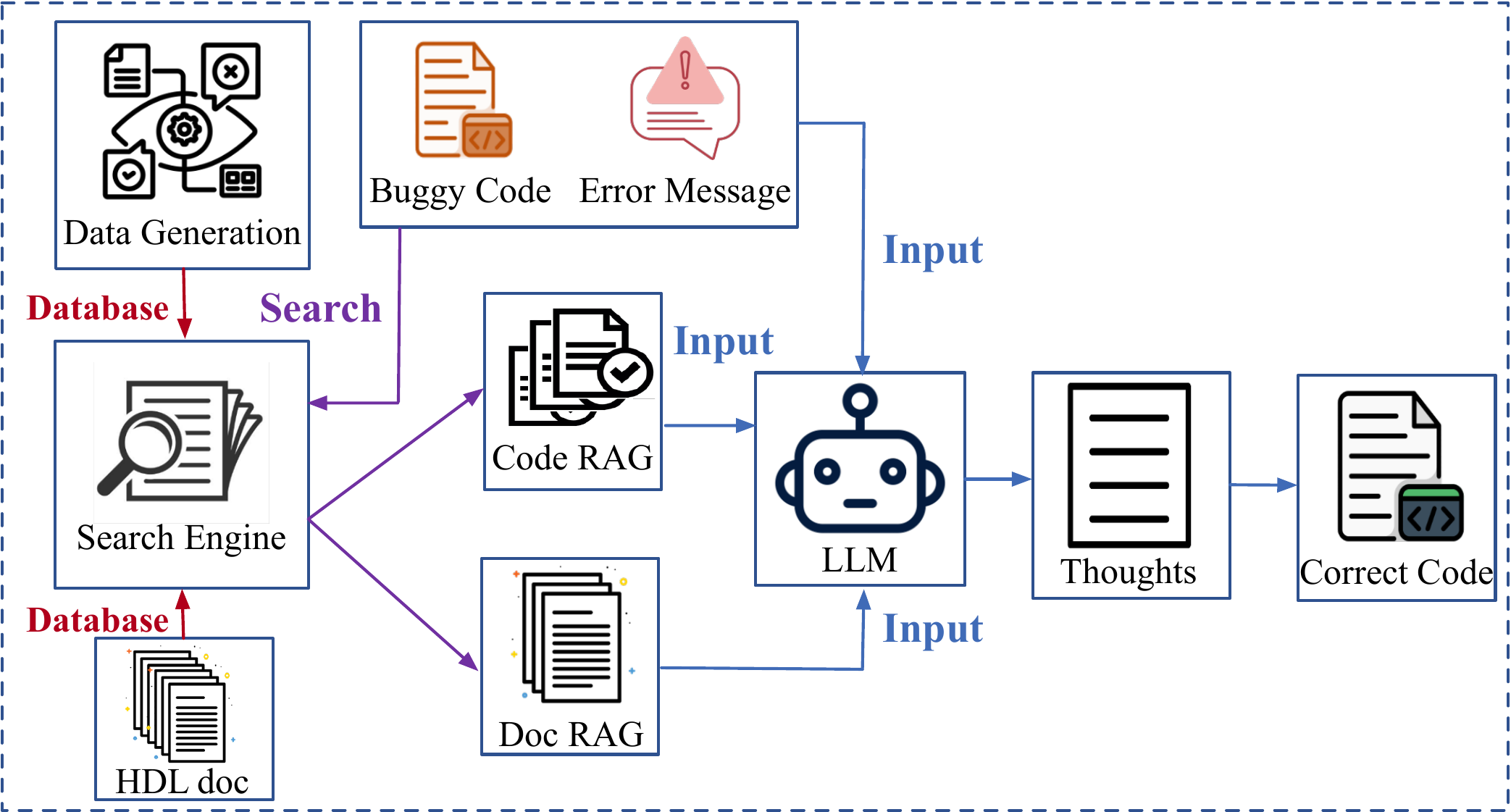}
    \caption{Framework overview of HDLdebugger.}
    \label{fig:framework}
\end{figure}

 \fakeparagraph{Framework}
As shown in Fig.~\ref{fig:framework},
given a buggy code and its associated error message, 
our proposed HDLdebugger targets to repair this buggy code into the correct one. 
Specifically, we first propose a data generation approach in Sec.~\ref{ssec:data-aug}  to generate a set of HDL code instances, where each instance consists of buggy code, error messages, and its correct version. 
These generated HDL code instances will be used to provide context for buggy code queries and fine-tune the LLMs.
Second, we propose a search engine in Sec.~\ref{ssec:search_engine}, which targets to retrieve relevant text information(i.e., document RAG) for error messages and retrieve buggy codes with similar buggy patterns (i.e., code RAG).  
Then, HDLdebugger takes the buggy code, its error message, task prompt, document RAG, and code  RAG to the LLMs, and enables LLMs to predict the correct code. Specifically, we introduce a retrieval-augmented fine-tune approach to fine-tune the LLMs for HDL code debugging in Sec.~\ref{ssec:LLM-train}.

\begin{figure}
    \centering
    \includegraphics[height=4.6cm]{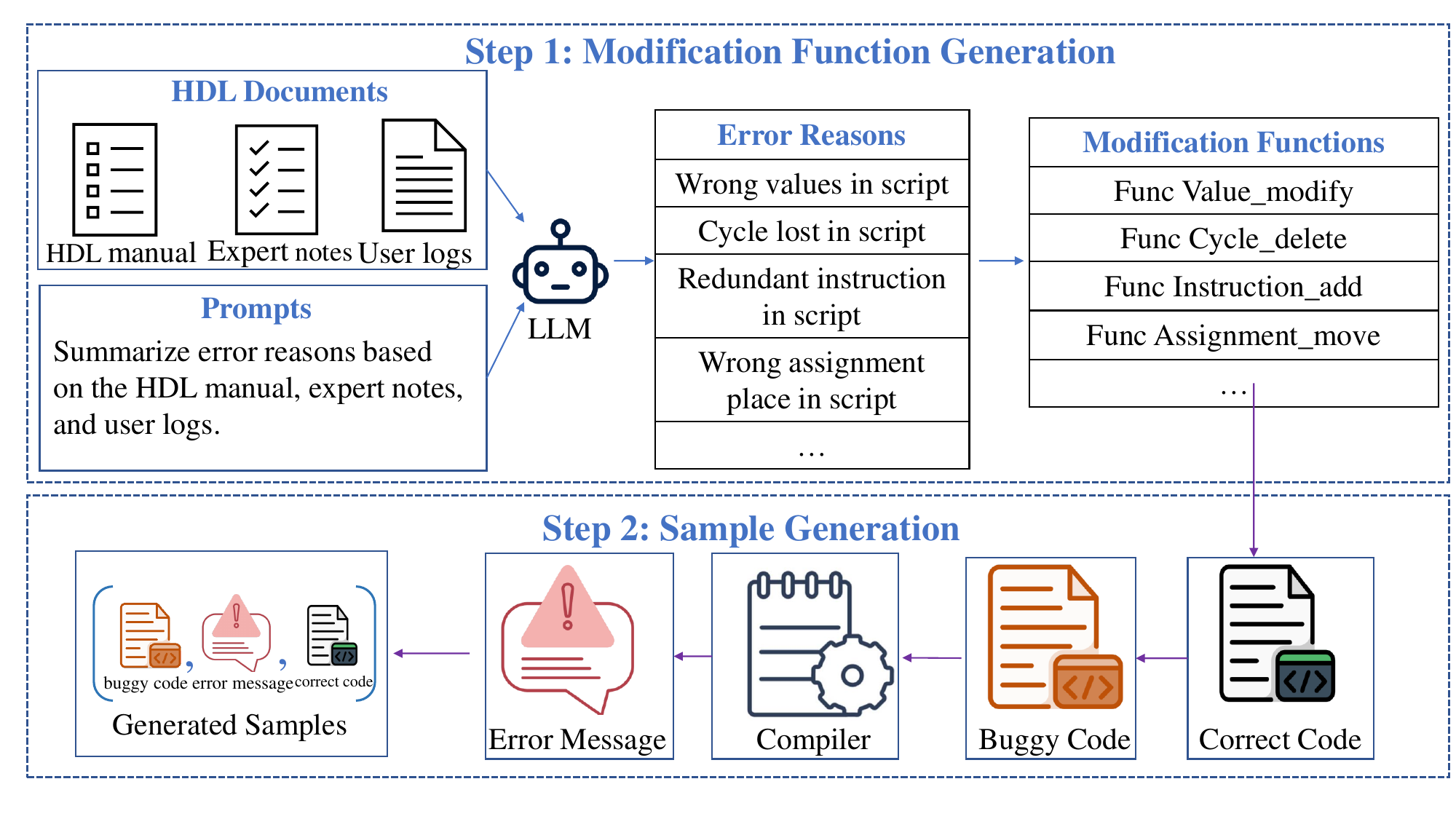}
    \caption{Reverse Engineering Pipeline for inducing error to correct code by diverse modification functions summarized from HDL documents.}
    \label{fig:reverse_engineer}
\end{figure}

\subsection{HDL Buggy Data Generation} \label{ssec:data-aug}

Unlike software languages like Python or C++, where code can often be crawled from open websites and platforms like GitHub, HDL codes, particularly those used for chip testing, are rarely made public due to privacy and commercial concerns.
This limitation presents a significant obstacle on fine-tuning LLMs in the domain of HDL.
In this section, we will introduce a reverse engineering pipeline for generating high-quality HDL code pairs that consist of both buggy and corrected versions. 
As shown in Fig.~\ref{fig:reverse_engineer}, The HDL data generation consists of two steps, i.e., modification function generation and sample generation.

\subsubsection{Modification Function Generation}
The first step targets to generate a set of high-quality modification functions.
These functions are then employed to modify HDL codes provided by industry engineers, thereby generating a diverse collection of buggy code examples.
To ensure that the modified codes exhibit a range of realistic and diverse error patterns, we construct the modification functions by leveraging the capabilities of LLMs and a comprehensive collection of industrial HDL documents.
Specifically, as shown in Fig.~\ref{fig:reverse_engineer},  we first collect a set of HDL documents, including the HDL manual, expert notes, and user logs. 
Subsequently, 
we then carefully design a prompt that guides the LLMs to extract and summarize prevalent and critical error patterns in HDL, such as syntax misuse or logical errors.
With these insights, we proceed to develop the modification functions.
The distilled functions focus on simple operations such as adding, deleting, modifying, and adjusting segments with HDL scripts, while the unique set of rules governing HDL ensures that similar operations can result in vastly different errors recorded in HDL documents.
These functions explicitly introduce errors into the original correct HDL codes that mirror those commonly encountered errors in industry, thereby creating an invaluable dataset to fine-tune LLMs.

\subsubsection{Sample Generation}
In this step, 
we gather a broad range of accurate and high-quality HDL codes $C=\{c_i\}_{i=1}^{|C|}$ from experienced chip engineers.
These codes, which have been utilized across various chip designs, are comprehensive to encompass a wide range of functional testing scenarios for chips.
These various HDL codes serve as the seed code for error case
construction. 
Specifically, 
by applying the previous modification functions to these correct HDL codes, we systematically introduce errors, thereby producing a set of buggy codes.
Particularly, we can apply various modification functions to one HDL code, which allows us to generate multiple instances of buggy code. 
Next, we employ an HDL compiler to compile these intentionally buggy codes on their respective design, which inevitably results in compilation errors. 
Then, for each correct HDL code $c_i \in C$, we can collect one of its buggy code $b_i$ with associated error message $m_i$ as an instance $I_i=(b_i,m_i,c_i)$ of training data.
Normally, diagnosing and rectifying HDL errors require the expertise of chip engineers. Given that our instructions systematically brought the errors, we can employ reverse engineering to identify the faults and produce solutions directly. This bypasses the need for manual error diagnosis, streaming the process of creating a vast array of comprehensive datasets of error scripts, error messages, and corresponding solutions.
This method ensures a rich diversity in the types of errors produced, which is critical for creating an extensive and effective training dataset $D_c=\{I_j\}_{j=1}^{|D_c|}$.

\subsection{Search Engine for RAG}\label{ssec:search_engine}
\begin{figure}[t]
    \centering
    \includegraphics[height=4.95cm]{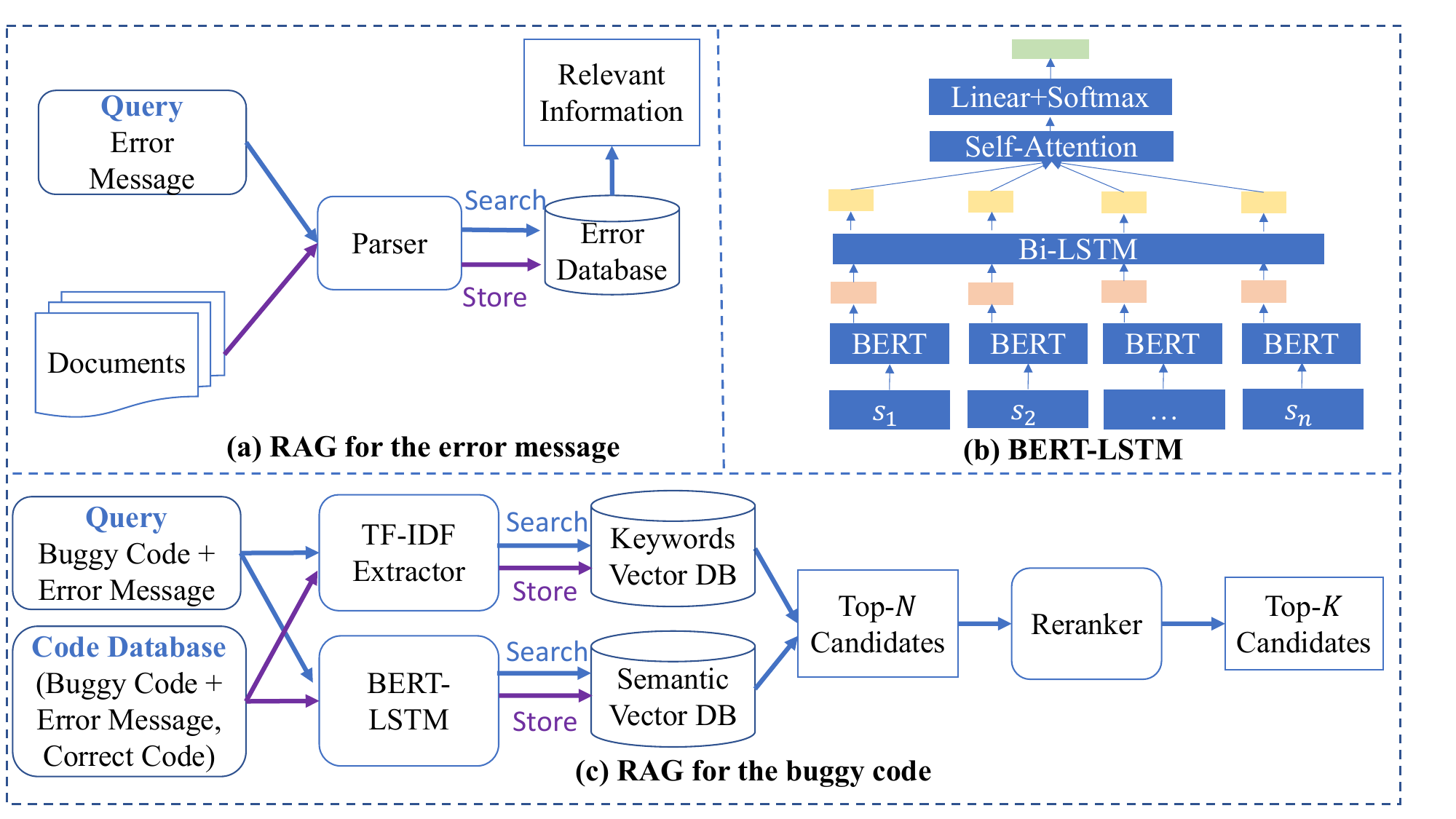}
    \caption{The search engine to retrieve relevant information given error messages and retrieve similar buggy codes based on a buggy code query.}
    \label{fig:method:RAG}
\end{figure}

In this subsection, we propose a search engine to optimize retrieval-augmented generations (RAG) for retrieving relevant information in the HDL documents and codes instances.
The retrieved RAG content will serve as contextual information for queries, thereby enhancing the capability of LLMs to understand buggy context information and identify issues within buggy codes.
We illustrate the overview of search engine framework in Fig.~\ref{fig:method:RAG}.


\subsubsection{Document RAG}\label{sssec:document_rag}
As shown in Fig.~\ref{fig:method:RAG}~(a),
we first collect a comprehensive collection of instructional documents for this HDL, encompassing language specifications, error diagnostics, and troubleshooting techniques.
Then,  we meticulously curate the content of documents, distilling a dedicated error database $D_e$ tailored to this HDL. This error database contains detailed error descriptions $d_j$, underlying reasons $r_j$, and suggested remedial strategies $s_j$ for each error id $e_j$.
{ We illustrate examples of error information in error database in Tab.~\ref{table:RAG-DRC_RULE} in the Appendix.}
Given an error message query $m_i$, we first parse $m_i$ to extract $n_{e,i}$ constituent error codes $\{e_j\}_{j=1}^{n_{e,i}}$. 
Subsequently, we retrieve the descriptions, reasons, and potential solutions for each identified error code from our error database.
Then, the document RAG of the query $m_i$ can be assembled as
$rag_i^d=\{(e_j,d_j,r_j, s_j)\}_{j=1}^{n_{e,i}}$, thereby helping LLMs to understand the error message $m_i$.

\subsubsection{Buggy Code RAG}\label{sssec:code_rag}
As shown in Fig.~\ref{fig:method:RAG}~(c),
in the code retrieval component, we maintain a code database $D_c=\{I_i\}_{i=1}^{|D_c|}$, where each code instance $I_i=(b_i,m_i,c_i)$ consists of a buggy code $b_i$, its associated error messages $m_i$, and the correct code $c_i$.
Given a query $q = (b, m)$ that includes a snippet of buggy code $b$ with its associated error message $e$, the aim of the code  RAG is to identify a subset $D_c^q \subseteq D_c$ of the top-$k$ code instances that have the most similar to the buggy code $b$ in the query.
In such a way, 
the buggy code  $b$ has similar buggy patterns with the buggy codes in each instance  $I_i=(b_i,m_i,c_i) \in D^q_c$.
Thus, the correct code $c_i$ for each buggy code $b_i$ can be provided to LLMs. 
As a result, LLMs can use
the learned patterns from the top-$k$ similar instances to
  fix for the buggy code $b$ in the query. 
  Specifically, we first introduce how to learn low-dimensional vectors for buggy codes and then then introduce a two-stage ranker that retrieves the top-$k$ buggy code instances for code query $q$.
  
 \fakeparagraph{Vector Database Construction}
 Given the complexity and length of HDL code and associated error messages, as well as similar-looking code snippets containing distinct buggy patterns, computing the similarity between buggy codes is challenging.
To address this challenge, we propose to measure the similarity between buggy from two aspects, i.e., keyword similarity and semantic similarity. 
First, we extract words from all buggy codes and their error messages and use the TF-IDF~\cite{aizawa2003information} technique to compute the weight of each word. Then, we can generate the keyword vector $\mathbf{z}^w_i$ for each buggy code $b_i$ with its error message $m_i$.
Second, we design a BERT-LSTM model that combines BERT, for its powerful language understanding capabilities, with an LSTM, for its sequential data processing strengths. 
The BERT-LSTM model encodes a buggy code $b_i$ with its error message $m_i$ into a low-dimensional embedding $\mathbf{z}^s_i$. 
Specifically, as introduced in Sec.~\ref{ssec:data-aug}, the buggy codes are generated by different modification functions, and here we take these modification functions as the labels for each buggy and optimize the BERT-LSTM model.
Then, we use the final representation of BERT-LSTM as the semantic embedding for each buggy code and its error messages.
The architecture details of BERT-LSTM models are introduced in Appx.~\ref{appx:bert_lstm}. 
Based on the TF-IDF and BERT-LSTM models, we build a keyword vector database and a semantic vector database, which together facilitate a robust framework for analyzing the similarity between instances of buggy HDL code.

 \fakeparagraph{Two-stage Ranker}
In general, we design a two-stage ranking approach to identify the top-$k$ most relevant buggy code instances to a given query of buggy code. 
In the first-ranking stage, for any given buggy code query $q=(b,e)$, we first use the TF-IDF encoder and BERT-LSTM encoder to generate the keyword vector $\mathbf{z}^w_i$ and semantic vector $\mathbf{z}^s_i$. 
Then, we define the similarity between the query code $q$ and each instance $I_i \in D_c$ in the code database $D_c$ as:
\begin{equation}\label{eq:sim_score}
    sim(q, I_i) = \lambda \cdot cosine(\mathbf{z}^w,\mathbf{z}^w_i) + (1-\lambda)\cdot cosine(\mathbf{z}^s,\mathbf{z}^s_i) +1,
\end{equation}
where $\lambda \in [0,1]$ is a hyper-parameter between semantic similarity and keyword similarity and $ sim(q, I_i) \in [0,2]$. 
We select the top-$N$ similar instances $\hat{D}^q_c$ from the code database $D_c$ based on the similarity score.
In the second-ranking stage, our goal is to pinpoint the top-$k$ relevant yet diverse buggy codes. These selections are aimed at providing a broader range of buggy patterns, which can help LLM repair the bugs in the query.
Formally, given $N$ buggy instances $\hat{D}^q_c$ and the query buggy code $q=(b,e)$, we select the top-$k$ relevant and diverse code $D^q_c$ by maximizing the following objective: 
\begin{align}\label{eq:rerank_obj}
    \max_{D^q_c \subseteq \hat{D}^q_c}{\sum_{I_i \in D^q_c}{sim(q,I_i)} +   \frac{1}{k}\cdot \sum_{I_i \in D^q_c}{dis(I_i, D^q_c)} },
\end{align}
where distance $dis(I_i, D^q_c)=\min_{I_j \in D^q_c \setminus I_i}{(2-sim(I_i,I_j))}$ denotes the diversity value between each instance $I_i$ and the other instances in $D^q_c$ and $dis(I_i, D^q_c) \in [0,2]$.
Eq.~\eqref{eq:rerank_obj} is an NP-hard problem, which can be reduced from the well-known $k$-clique problem~\cite{tsourakakis2015k} by setting $sim(q,I_i)=1$ for all $I_i \in D_c$.
Therefore, we propose a greedy algorithm with an approximation ratio $1-1/e$ to identify the top-$k$ relevant buggy codes.
The details of the greedy algorithm and approximation ratio are introduced in Appx.~\ref{appx:greedy_algorithm}.
Thus, given a buggy code $b_i$ and error message $m_i$, we can generate the code retrieval-augmented generation $rag^c_i=\{(b_j,m_j,c_j)\}_{j=1}^{k}$.

\subsection{Retrieval-augmented LLM Fine-tuning} \label{ssec:LLM-train}
In this subsection, we introduce how to fine-tune the LLMs based on the training dataset constructed in Sec.~\ref{ssec:data-aug} and the search engine proposed in Sec.~\ref{ssec:search_engine}.
Specifically, we first propose to generate a thought to help LLM repair each buggy code in the training dataset. 
Second, based on the generated thought and the retrieved buggy codes, we propose a retrieval-augmented supervised fine-tuning technique for LLMs. 

\subsubsection{Self-guided Thought Generation}
One straightforward way is to feed buggy code and error messages to LLMs and let LLMs directly predict the correct codes. 
However, this approach is insufficient for LLMs to deeply comprehend the problem and provide accurate solutions. 
Recent research~\cite{wei2022chain,yao2023tree} suggests that when LLMs are prompted to produce a series of intermediate and explanatory thought before finally outputting the solution to the given task, the performance of LLMs can be significantly improved.
It is because these reasoning thoughts improve the understanding of LLMs on the input tasks and thus generate more relevant and accurate outputs.
Therefore, before fine-tuning the LLMs, we propose to generate high-quality thought for each training code instance.

\begin{figure}[th!]
    \centering    
    \includegraphics[height=12.3cm]{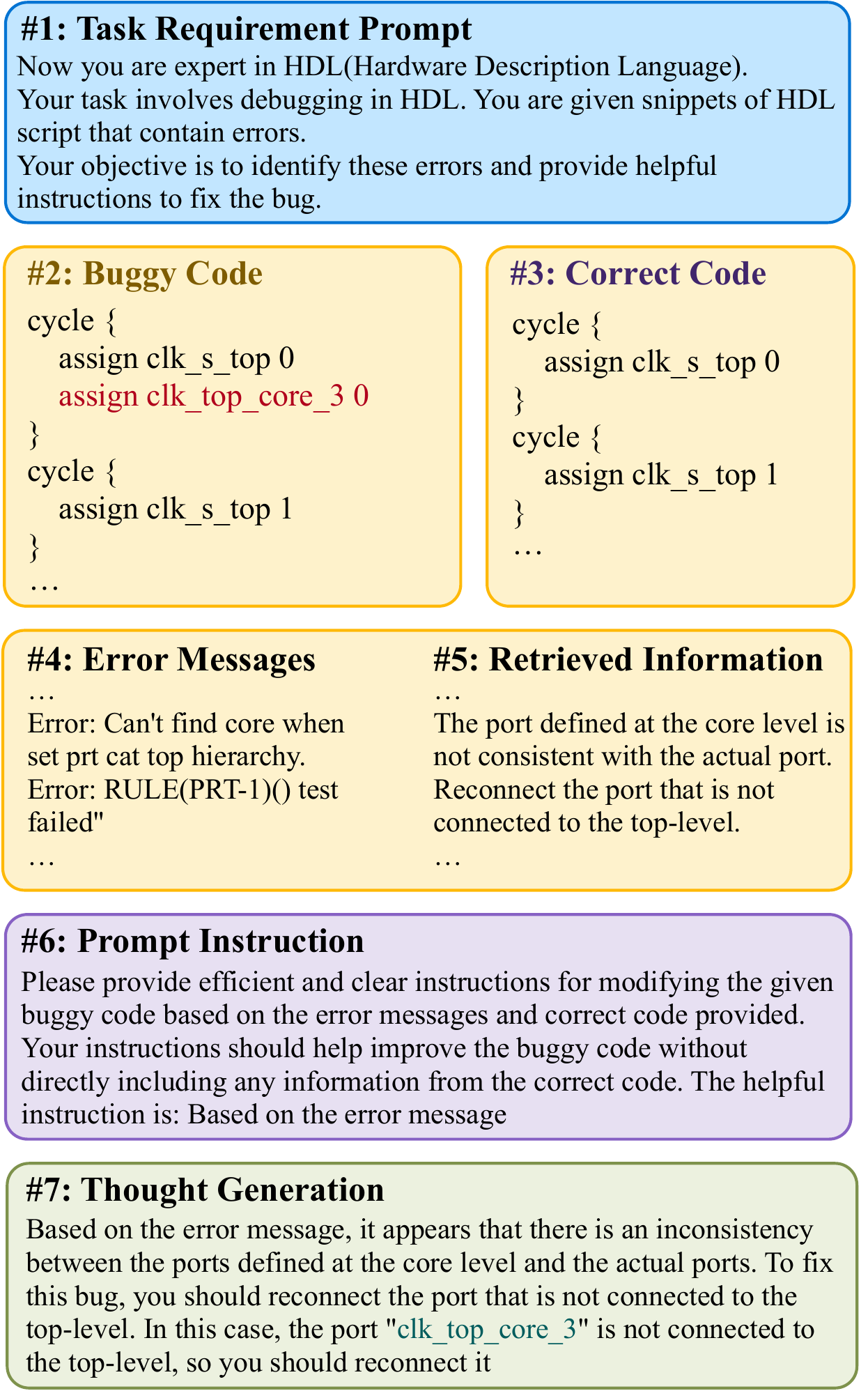}
    \caption{Thought generation example.}
    \label{fig:method:thoughts-generation}
\end{figure}

Specifically, as illustrated in Fig.~\ref{fig:method:thoughts-generation}, we first design a precise and explicit prompt $p_t$ to clarify the thought generation task for LLMs. 
Following this, we input the thought generation prompt $p_t$, the buggy code $b_i$, its associated error message $m_i$, and the document RAG $rag_i^d$, and its correct version $c_i$ into the LLM with its inference mode $LLM_1$. This enables the LLM to generate a thought $t_i$ on how to repair the buggy code $b_i$ into the correct code $c_i$ as follows:
\begin{equation}
    \label{eq:thoughts-generation}
    t_i = LLM_1(p_t \circ b_i \circ m_i \circ c_i \circ rag_i^d ),
\end{equation}
where $\circ$ means concatenate operator. 
We omit the general task requirements prompt for simplification.
Empirically, we find that when LLMs generate the thought $t_i$ only once, the output might be irrelevant to the buggy code $b_i$ or incorrect due to the hallucination phenomenon and randomness of LLMs.
Therefore, to guarantee the quality of the generated thought, we iteratively generate $L$ different thoughts $T_i = \{t_{i,j}\}_{j=1}^L$ for each buggy code $b_i$ by running  $LLM_1$ $L$ times following Eq.~\eqref{eq:thoughts-generation}. 
This is achieved by modifying temperature parameters and the details are described in appendix.

To assess the quality of each thought $t_{i,j} \in T_i$, we adopt a self-guidance strategy to select the highest quality thought from the thought set $T_i$ for the buggy code $b_i$. 
Specifically, we feed the buggy code $b_i$, its associated error message $m_i$, and the document RAG $rag_i^d$ 
to the LLM.
A prompt $p_c$, "Based on the analysis, the correct script is," is appended to guide the LLM towards generating the predicted correct script, represented as:
\begin{equation}
    \hat{c}_{i,j} = LLM_1(p_c \circ b_i  \circ m_i \circ rag_i^d \circ t_{i,j}).
\end{equation}
After we obtain the output $\hat{c}_{i,j}$ for each thought $t_{i,j}$, we employ the edit distance metric~\cite{navarro2001guided}to evaluate the similarity between the predicted correct code $\hat{c}_{i,j}$ and  the ground-truth of corrected code $c_{i}$ as
\begin{equation}
    d_{i,j} = EditDistance(c_i, \hat{c}_{i,j}).
\end{equation}
Intuitively, if the distance $d_{i,j}$ between the predicted correct code $\hat{c}_{i,j}$ and the ground truth of corrected code $c_{i}$ is smaller, the thought $t_{i,j}$ is more helpful to LLMs to repair the buggy code $b_i$.
Therefore, we select the thought with the smallest edit distances for each $b_i$, i.e., $t_i = \min_{t_{i,j} \in T_i}{d_{i,j}}$.
Finally,  we can obtain the training dataset $\mathcal{D} = \{(b_i, m_i, rag_i^d, rag^c_i, t_i, c_i)\}_{i=1}^{|D_c|}$.
The details of the thought generation are illustrated in Alg.~\ref{alg:method:thoughts-generation} in the Appendix.

\subsubsection{Retrieval-augmented Fine-tuning}
After obtaining the final training dataset $\mathcal{D} = \{(b_i, m_i, rag_i^d, rag^c_i, t_i, c_i)\}_{i=1}^{|D_c|}$, we supervise fine-tune the LLMs based on buggy codes and retrieval-augmented generation in Sec.~\ref{ssec:search_engine}.
Specifically, given each training instance $D_i=(b_i, m_i, rag_i^d, rag^c_i, t_i, c_i)\in\mathcal{D}$, we first feed thought generation prompt $p_t$, buggy code $b_i$, its error message $m_i$, document RAG $rag_i^d$, code RAG $rag^c_i$, to a target LLM, $LLM_2$ to generate the predicted thought $\hat{t}_i$ as follows:
\begin{equation}
    \label{eq:thought_pred}
    \hat{t}_i = LLM_2(p_t \circ p_c \circ b_i \circ m_i  \circ rag_i^d \circ rag^c_i).
\end{equation}
Then, based on the predicted thought $\hat{p}_t$ and code correction prompt $p_c$, LLM $LLM_2$ predicts the correct code $\hat{c}_i$ as follows:
\begin{equation}
    \label{eq:code_pred}
    \hat{c}_i = LLM_2(p_t \circ p_c \circ b_i \circ m_i  \circ rag_i^d \circ rag^c_i \circ \hat{t}_i).
\end{equation}
Following~\cite{wang2022self,alpaca}, 
we fine-tune the target LLM with its training mode $LLM_2$ by using the conventional next-token prediction objective and minimize the cross-entropy loss $ \mathcal{L}_{\mathcal{D}}$:
\begin{align} 
    & \mathcal{L}_{{t}_i} = \sum_{w_{i,j} \in t_i}\log p_{LLM_2}(w_{i,j}|C \circ w_{i,<j}), \nonumber \\
    & \mathcal{L}_{{c}_i} = \sum_{w_{i,k} \in c_i}\log p_{LLM_2}( w_{i,k}|C \circ \hat{t}_i \circ w_{i,<k}), \nonumber\\
    & \mathcal{L}_{\mathcal{D}} = \frac{1}{2|\mathcal{D}|}\sum_{D_i \in \mathcal{D}}{(\mathcal{L}_{{t}_i} + \mathcal{L}_{{c}_i})},
\end{align}
where the context $C=p_t \circ p_c \circ b_i \circ m_i \circ rag_i^d \circ rag^c_i$ denotes the concatenate of the inputs for clarification $w_{i,j} \in t_i$ denotes the $j$-th word in $t_i$ and $w_{i,<j}$ denotes a set of words in $t_i$ before $w_{i,j}$, and $ p_{LLM_2}(w_{i,j}|w_{i,<j})$ denote the probability of $w_{i,j}$. $\mathcal{L}_{{t}_i}$ and $\mathcal{L}_{{c}_i}$ denote the prediction cross-entropy loss on thought ground truth $t_i$ and correct code ground truth $c_i$ regarding the buggy code $b_i$, respectively.


\section{Experiments}\label{sec:exp}

\subsection{Experiment Setting} \label{exp:setup}
\subsubsection{HDL Datasets in Huawei}\label{sssec:datasets}
We gather a diverse collection of HDL code files from Huawei.
These files are meticulously curated, with each file being specifically utilized for distinct chip design scenarios, reflecting the varied and specialized demands of circuit design.
Each HDL code file contains an extensive array of data, including variable assignments, detailed circuit designs, clocking information, function testing protocols, and various test items that are critical for the chip design process.
Based on these HDL code files, we use the data generation in Sec.~\ref{ssec:data-aug} to generate 92,143 distinct HDL training code instances. 
Each HDL training code instance consists of the buggy HDL code, error messages, and the correct HDL code.
Specifically, in the experiments, we split the data into training and testing sets at a ratio of 8:2, respectively.

\subsubsection{Baselines}
We compare our proposed HDLdebugger with 13 baselines in three types of code debugging approaches as follows:
\begin{itemize}[leftmargin=12pt]
 \item \textbf{Five large language models}: We compare HDLdebugger against two profound and state-of-the-art LLMs available through API services, including ChatGPT~\cite{brown2020language} and GPT-4~\cite{achiam2023gpt}. Additionally, we compare HDLdebugger with three open-source LLMs: OpenChat~\cite{wang2023openchat}, Orca2~\cite{mitra2023Orca}, and Mistral~\cite{jiang2023Mistral}. These open-source models have shown performance on par with ChatGPT across various open LLM benchmarks.

 \item \textbf{Four Code Debugging and HDL Models}: We compare HDLdebugger with two code debugging and hardware code generation models, i.e., Self-debug~\cite{chen2023teaching} and RTLfixer~\cite{tsai2023rtlfixer}.
Self-debug~\cite{chen2023teaching} is one of the most classical methods of code debugging.
RTLfixer~\cite{tsai2023rtlfixer} is proposed to solve HDL debugging problems. 
VeriGen~\cite{thakur2023verigen} and RTLCoder~\cite{liu2023rtlcoder} are two LLMs targeting hardware language.

\item \textbf{Four Code Language Models}:
We compare four SOTA pretrained code LLMs, i.e., Deepseek~\cite{bi2024deepseek}, Starcoder~\cite{li2023starcoder}, Stablecode~\cite{rombach2022high}, and WizardCoder~\cite{luo2023wizardcoder}.
\end{itemize}
For baselines except for ChatGPT~\cite{brown2020language} and GPT-4~\cite{achiam2023gpt}., we adopt three strategies, i.e., the raw model, the raw model with RAG, and the raw model with supervised fine-tuning (SFT).
\begin{itemize}[leftmargin=12pt]
 \item \textbf{Raw model}: We only feed buggy code and error messages to the model and enable raw models to infer the correct code directly. 
 
 \item \textbf{Raw Model with RAG}: For buggy code, we feed the buggy code, its error message, and the document RAG and code RAG obtained in Sec.~\ref{ssec:search_engine} to the raw model and enable the raw models to infer the correct code directly.

 \item \textbf{Raw Model with SFT}:
 We take the buggy code and error message as inputs of LLMs and use the correct code as ground-truth to fine-tune the raw models.
\end{itemize}
Specifically, 
we selected CodeLlama-13b as our base model from the available code LLMs. 
The choice of 13b was driven by its optimal model size, which strikes a balance between training and deployment, taking into account both performance and cost factors.

\subsubsection{Evaluation Metrics}\label{sssec:metrics}
For the overall debug system, we mainly evaluate its pass rate for correcting codes, relative code runtimes, and edit distance between correct code and buggy code. The calculation of these metrics is listed below.
\begin{itemize}[leftmargin=8pt]
    \item \textbf{Pass-Rate}: Pass rate for executing code file corrected by each method is defined as $P=\frac{\sum_{i=1}^{n_c}\mathbb{S}(y_i)}{n_c}$, where $y_i$ stands for the corrected code, $\mathbb{S}$ denotes for executing code successfully.

    \item \textbf{Run-Time}: The relatively average compilation time for Huawei's internal HDL compiler to execute all test code files. The Run-Time for results from our HDLdebugger is set as the base unit.
    
    \item \textbf{Edit-Distance}:
    Edit-Distance calculates the minimum number of operations (insertion, deletion, substitution) required to transform one code snippet into the other. 
    
\end{itemize}

Also, we evaluate our code search engine in Sec.~\ref{ssec:search_engine} by the hit ratio, mean average precision, and mean reciprocal rank metrics, which are formulated as follows.
\begin{itemize}[leftmargin=8pt]
    \item \textbf{H@K}: Hit ratio for top-$K$ recommendation results on $n_t$ code queries is formulated as $H@K=\frac{1}{n_t}\sum_{i=1}^{n_t}\frac{1}{K}\sum_{k=1}^{K}\mathbb{I}(y_{i,k},y_{i})$, where $y_{i,k}$ denotes the error type of the retrieved $k$-th buggy code for query code $b_i$, and the indicator function $\mathbb{I}(y_{i,k},y_{i})=1$ if $y_{i,k}=y_{i}$
    
    \item \textbf{MAP@K}: Mean average precision (MAP) for top-$K$ results on $n_t$  queries is defined as $MAP@K=\frac{1}{n_t}\sum_{i=1}^{n_t}\frac{1}{K}\sum_{k=1}^{K}\frac{\mathbb{I}(y_{i,k},y_{i})\cdot n(b_{i,\le k})}{k}$, where $n(b_{i,\le k})$ denotes the number of recommendations in the first top-$k$ that has the same error label with query $b_i$.
    
    \item \textbf{MRR@K}: Mean reciprocal rank (MRR) for top-$K$ results on $n_t$ queries is formulated as $MRR@K=\frac{1}{n_t}\sum_{i=1}^{n_t}\frac{1}{K}\sum_{k=1}^{K}\frac{\mathbb{I}(y_{i,k},y_{i})}{k}$.
\end{itemize}

\begin{table}[tb!]
    \small
    \caption{Main Results. Pass-Rate describes the absolute value, which is the higher the better. $< 1\%$ describes the Pass-Rate less than 1\%. Run-Time and Edit-Distance are relative values compared with ours, which are both the lower the better. }
    \begin{tabular}{c|c|c|c}
    \toprule
                     Method & Pass-Rate & Run-Time &  Edit-Distance \\ \midrule
    ChatGPT$^*$         &    3.01\%            &    2.25 &      31.28\\
    GPT4$^*$         &     6.35\%           &  1.94   &     18.17 \\
    OpenChat &     <1\%           &  2.51   &                   34.47        \\ 
    OpenChat  w/ RAG &   3.01\%             &   2.21  &                       10.37    \\ 
    Orca2 &      2.68\%          & 2.08    &                       2.94    \\ 
    Orca2 w/ RAG &     9.03\%           &  2.10   &                  4.74         \\ 
    Mistral &      7.36\%          &   2.39  &                    6.68       \\ 
    Mistral w/ RAG &     24.75\%           &   2.01  &                      9.88     \\
    \midrule
    Self-debug       &   5.02\%             & 2.49    &                10.24           \\
    RTLfixer       &     28.35\%           &  2.11   &                         11.05  \\
    VeriGen      &         <1\%        &   2.46  &     49.16                      \\
    VeriGen w/ RAG      &   1.34\%             &   2.56  &                        37.12   \\
    VeriGen  w/ SFT     &    67.55\%            &  1.35   &                 1.38          \\
    RTLCoder        &        <1\%         & 2.49    &    43.08                       \\
    RTLCoder  w/ RAG       &    <1\%              &  2.65   & 34.85                          \\
    RTLCoder   w/ SFT      &  64.21\%                &  1.53    &               3.83         \\
    \midrule
    Deepseek       &       2.34\%              &      2.58    &                                  32.17       \\
    Deepseek  w/ RAG     &  3.34\%                   &  2.51        &                              34.38           \\
    Deepseek  w/ SFT     &   51.63\%                  &  1.64        &                           3.35              \\
    Starcoder       &          <1\%           &  2.58        &                                   28.85      \\
    Starcoder  w/ RAG     &    <1\%                  &    2.56      &                                 34.85        \\
    Starcoder  w/ SFT     &    68.27\%                 &  1.21        &                              1.57           \\
    Stablecode  &          6.69\%           &  2.46        &                         4.56     \\
    Stablecode w/ RAG &     8.02\%                &     2.50     &                          6.25    \\
    Stablecode w/ SFT &    41.47\%                 &    2.38      &                   6.10      \\
    WizardCoder    &      3.01\%               &    2.41      &     7.19                      \\  
    WizardCoder w/ RAG         &   4.68\%                  &   2.58       &                8.67            \\
    WizardCoder w/ SFT        &   71.57\%                  & 1.04          &   1.06                        \\ \midrule
    HDLdebugger(ours)     &   \textbf{81.93\%}                  &    \textbf{1.00}      &      \textbf{1.00}                   \\ \bottomrule
    \end{tabular}
    \label{tab:main-results}
\end{table}

\subsection{Main Results}
~\Cref{tab:main-results} demonstrates the main performance of our results and other methods.
It's clear that our method outperforms other methods by all means by a large margin including direct approach, RAG and SFT, which demonstrates the effectiveness of our framework.
For both runtime and edit distance metrics, we normalize all results and present only the relative values in comparison with ours to enhance the clarity and effectiveness of the comparison.
\subsubsection{Comparison with different types of LLMs.}
In our study, we conducted a comparative evaluation of both general-purpose LLMs such as ChatGPT, GPT-4, OpenChat, Orca, Mistral, Deepseek, Starcoder, Stablecode, WizardCoder, and specialized code LMs including Self-debug, RTLfixer, VeriGen, RTLCoder, within the HDL debugging scenario. 
Due to privacy concerns, certain code specifications have been omitted for testing purposes when using ChatGPT or GPT4. To distinguish these versions, we will refer to them as ChatGPT$^*$ and GPT4$^*$.
It is evident that our approach exhibits superior performance against all 13 state-of-the-art benchmarks, including GPT-4 and other domain-specific hardware-based language models.

\subsubsection{Analysis of Different Strategies.}

In this study, we implement three distinct evaluation strategies to assess the efficacy of various methodologies: with retrieval-augmented generation (RAG), with supervised fine-tuning (SFT), and via a direct approach. Within the context of HDL debugging, our analysis reveals that SFT holds greater significance and applicability across all evaluated baselines, including VeriGen, RTLCoder, Deepseek, Starcoder, Stablecode, and WizardCoder. In the majority of scenarios, we note that methods enhanced through SFT consistently outperform those augmented with RAG by a substantial margin. Besides, our proposed method integrates both RAG and SFT strategies, achieving unparalleled performance, indicating that it is better to incorporate SFT and RAG for the HDL debugging task.

\vspace{-1em}
\subsubsection{Impact on Domain-Specific Solutions.}
In addition to general and code-specific Language Models (LLMs), methodologies such as Self-debug and RTLfixer are specifically devised to address code debugging scenarios. While these approaches demonstrate improvements over other general and code LLMs, their effectiveness still falls short of being fully satisfactory. Our analysis extends to evaluating our dataset with LLMs exclusively trained on hardware languages, namely VeriGen and RTLCoder. Contrary to expectations, these specialized hardware language models do not outperform their general and code LLM counterparts in our HDL debugging context, suggesting the possibility of an inherent task domain generalization issue within HDL debugging scenarios. 
On the other hand, our approach consistently surpasses domain-specific solutions regardless of the varied prompt engineering techniques employed or the domain-specific data used for training, which underscores the effectiveness of our methodology.

\vspace{-0.5em}
\subsection{Ablation Studies}

Firstly, we provide a detailed analysis of the impact of different strategies of our method.
~\Cref{tab:abl-main-results} illustrates the performance of different strategies. For baseline ,we only use direct inference strategy on base model, i.e., CodeLlama.
SFT w/ th indicates supervised fine-tuning with generated thoughts.
In terms of RAG \& SFT w/ th, we mainly refer to retrieval augmented LLM fine-tuning where both retrieved code instances and relevant information are combined together for LLM fine-tuning.
From~\Cref{tab:abl-main-results} we can observe that SFT w/ th outperforms baseline by a large margin. 
Moreover, combing RAG $\&$ SFT also significantly improves the performance.

\begin{table}[tb!]
    \small
    \caption{Ablation on different strategies}
    \begin{tabular}{c|c|c|c}
    \toprule
                     Method & Pass-Rate & Run-Time &  Edit-Distance \\ \midrule
    Direct(CodeLlama)         &       4.01\%              &    2.26      &                     35.51      \\
     + RAG         &   15.05\%                  &    2.20      &                      5.34     \\ 
    + SFT        &  70.56\%                   &    1.19      &                      1.19     \\  
    + SFT w/ th &    74.91\%                &   1.09       &                          1.13    \\
    +  RAG \& SFT w/ th &  \textbf{81.93\%}                 &    \textbf{1.00}      &                      \textbf{1.00}        \\

    \bottomrule
    \end{tabular}
    \label{tab:abl-main-results}
\end{table}

Besides, we evaluate our search engine for RAG, comparing it with four methods: F-IDF~\cite{aizawa2003information}, BM25~\cite{robertson2004simple}, random forest~\cite{rigatti2017random}, and XGBoost~\cite{chen2016xgboost}. 
TF-IDF and BM25 assess relevance scores between buggy codes, while Random Forest and XGBoost generate low-dimensional representations for codes and error messages, similar to our BERT-LSTM model's approach to computing relevance through representation similarity. 

~\Cref{tab:search_engine} indicates our engine surpasses all baselines in accurately retrieving and correcting buggy code queries, highlighting its superior ability to decode complex buggy code patterns beyond the capabilities of traditional and machine learning models. Traditional methods like TF-IDF and BM25 lack the depth to understand complex code bugs, while models like XGBoost and Random Forest fall short in semantic comprehension. Our engine effectively combines textual and semantic analysis, enhancing bug detection and correction.

\begin{table}[tb!]
    \footnotesize
\caption{Evaluation on the search engine in Sec.~\ref{ssec:search_engine} on the top-$k$ recommendation. 
XGB and RF denote the XFBoost and Random Forest, respectively.
MAP$@$1 and MRR$@$1 are the same as H$@$1 mathematically.
The first row \textit{Optimal} is the optimal performance under each metric. 
}
\begin{tabular}{c|ccc|cc|cc}
\hline
 & H@1 & H@3 & H@10 & MAP@3 &MAP@10 & MRR@3 & MRR@10 \\ \hline
\textit{Optimal} & \textit{1.00}    &  \textit{1.00}   &  \textit{1.00}    & \textit{1.00} &  \textit{1.00}    &   \textit{0.61}    &   \textit{0.29}     \\ \hline

TF-IDF & 0.97    &  0.87   &  0.75    & 0.86 &  0.71    &   0.55    &   0.24     \\ 
 BM25 & 0.95   & 0.80    &   0.59   & 0.81 &   0.54   &   0.53    &    0.22    \\  
XGB &    1.00 &  0.35   &  0.18    & 0.34   & 0.14  &   0.34    & 0.12       \\ 
 RF &   0.31  &  0.43   &  0.41   &   0.34   & 0.29 &   0.25    &  0.11      \\ \hline
 Ours &  \textbf{1.00}   & \textbf{0.98}    &  \textbf{0.94}    &  \textbf{0.98}  & \textbf{0.93}   &   \textbf{0.60}    &   \textbf{0.28}     \\ \hline
\end{tabular}
\label{tab:search_engine}
\end{table}

\subsection{Parameter Sensitivity}\label{ssec:param_sensitivity}
In the following experiments, we evaluate parameter sensitivity across different hyper-parameters including retrieved code instances samples and related inference time.
We also consider temperature and pass-rate$@k$ for various $k$.

\begin{figure}[tb!]
    \centering
    \begin{minipage}{0.48\linewidth}
        \centering
        \includegraphics[width=\linewidth,height=2.88cm]{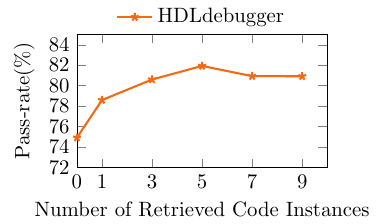}  
        \caption{Pass-rate and code retrieved code instances}
        \label{fig:exp-icl-sensitivity}
    \end{minipage}
    \hfill 
    \begin{minipage}{0.48\linewidth}
        \centering
        \includegraphics[width=\linewidth,height=2.88cm]{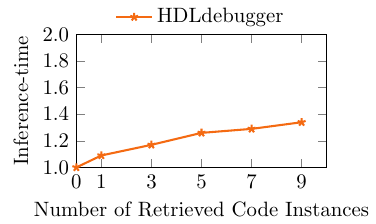} 
        \caption{Inference time and retrieved code instances}
        \label{fig:exp-icl-inference}
    \end{minipage}
\end{figure}

\begin{figure}[tb!]
    \centering    
    \includegraphics[width=0.408\textwidth]{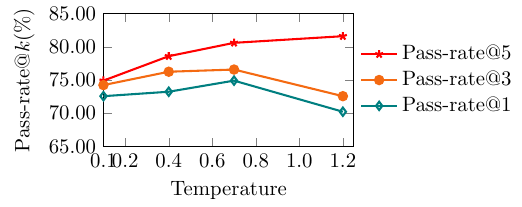}
	\caption{Pass-rate$@$k and Temperature}
	\label{fig:exp-pass-rate}
\end{figure}

\subsubsection{The number of retrieved code instances}
\Cref{fig:exp-icl-sensitivity} depicts the impact of varying the number of retrieved code instances on model performance.
We find that performance improves with each additional instance between 1 to 5, achieving the most significant gains. However, beyond five instances, gains plateau or even decrease, indicating diminishing returns. This suggests that while adding retrieved code instances enhances model performance up to a point, increasing instances beyond this threshold leads to inefficiency.
Moreover, we observe a tendency for the language model to generate repetitive or redundant content derived from previous input as the number of retrieved code instances increases. This phenomenon underscores a critical area for future exploration.

\subsubsection{Feasibility on Inference}
We assess the impact of incorporating the retrieved code instances on the additional inference budget. ~\Cref{fig:exp-icl-inference} illustrates the normalized inference times to clearly highlight the incremental budget required. A value of $0$ indicates the absence of retrieved code instances. Our observations reveal that as the number of retrieved code instances increases, the corresponding inference time exhibits a slow, logarithmic increase rather than a linear one. This pattern underscores the efficiency of our approach, demonstrating that integrating retrieved code instances significantly enhances performance without proportionally increasing the inference overhead.

\subsubsection{Pass-rate$@$k and Temperature}
We explore the effects of varying the temperature settings and the pass-rate$@$k for different values of $k$, where $k$ represents the number of answers generated by the LLM, as discussed in~\cite{chen2021evaluating}. 
Typically, a lower temperature setting yields more deterministic outcomes, whereas higher temperatures result in more varied outputs. ~\Cref{fig:exp-pass-rate} demonstrates that as $k$ increases, so does overall performance. 
Specifically, at a lower temperature, such as $0.1$, outputs are more consistent, leading to a narrower performance range. 
Conversely, at higher temperatures, like $1.2$, outputs become more varied, enhancing the likelihood of generating correct answers as $k$ increases. 
Notably, for a pass-rate$@$5, the performance at a temperature of $1.2$ surpasses that at $0.7$, indicating that increased temperature settings can improve outcomes, particularly at higher values of $k$.

\section{Related Work}\label{sec:related}

\subsection{Automatic Code Debugging}\label{ssec:pr}
Automatic code debugging has emerged as a promising area within software engineering~\cite{monperrus2018automatic,huang2023survey}.
Given a code with bugs, the task is to automatically fix the code bugs with the correct functions, which alleviates the burden of manual debugging and fixing code faults. Classic techniques can be mainly classified as template-based~\cite{jiang2018shaping,liu2019tbar}, heuristic-based~\cite{wen2018context,yuan2018arja}, constraint-based~\cite{xuan2016nopol,xiong2017precise}, and neural network-based approaches~\cite{fu2022vulrepair,chen2022neural,zhang2021autotrainer}.
Specifically, template-based approaches apply expert-defined code patterns to fix bugs. These approaches can only repair codes in specific patterns and lack generalization to other bugs. 
Heuristic-based approaches apply predefined heuristics and cannot cover all types of bugs.
Constraint-based approaches repair buggy codes by solving a constraint problem. These methods can be accurate, but they are computationally expensive.
Neural network-based approaches need numerous high-quality labeled training data pairs (i.e., pairs of buggy codes and fixed codes) to optimize parameters, which is time-consuming to collect the high-quality code pairs.

Recently, large language models (LLMs) have shed new light on automatic code debugging. 
The prevailing hypothesis suggests that LLMs, through training on extensive repositories of open-source code snippets, are adept at identifying bug patterns and facilitating the repair of defective code. 
Contemporary strategies employing LLMs predominantly utilize retrieval augmented generation (RAG) and sophisticated prompt engineering techniques to address debugging challenges. 
Notably, Self-debug~\cite{chen2023teaching} represents a pioneering effort in applying LLMs to code debugging, employing targeted prompt engineering for enhanced effectiveness.
RTLfixer~\cite{tsai2023rtlfixer} utilizes both RAG and prompt engineering to tackle the HDL debugging problem.
However, these methods do not show satisfactory results in our industry-level cases due to the lack of requisite knowledge of HDL codes.
Fine-tuning with HDL code resources is one alternative to tackle the problem.
Nevertheless, these approaches need abundant and high-quality labeled data, which is not suitable for HDL codes, since the related HDL codes are limited due to privacy and commercial issues.

\vspace{-0.5em}
\subsection{large language models for Code Generation}\label{ssec:llm}
Large language models (LLMs) have transformed the landscape of code generation by leveraging vast amounts of code data to predict and generate syntactically and semantically correct code snippets~\cite{wang2023software}. Notable among these models is OpenAI's Codex~\cite{chen2021evaluating}, which powers GitHub Copilot, offering context-aware code suggestions and completions to developers directly within their IDEs.
Another key contribution is from DeepMind's AlphaCode~\cite{li2022competition}, which excels in generating code solutions for competitive programming challenges and obtains the top percentile of participants in coding competitions. 
Recent advancements in LLMs tailored for the coding domain have seen significant contributions, with notable examples including DeepSeek~\cite{bi2024deepseek}, Starcoder~\cite{li2023starcoder}, Stabelcode~\cite{rombach2022high}, Codellama~\cite{roziere2023code}, and WizardCoder~\cite{luo2023wizardcoder}. These models have been trained on extensive datasets, both proprietary and open-source, to enhance capabilities in code generation, completion, and debugging. Within the realm of hardware description languages (HDLs), ChipNeMo~\cite{liu2023chipnemo} represents a pioneering effort in developing a domain-specific LLM, highlighting the challenges and potential of applying LLMs to the hardware domain. Despite being trained on vast hardware-specific datasets, ChipNeMo achieves performance that, while competitive, does not surpass that of state-of-the-art general LLMs, such as GPT-4. This underscores the inherent complexities of adapting LLMs to the nuances of HDL. Concurrently, initiatives like VeriGen~\cite{thakur2023verigen} and RTLCoder~\cite{liu2023rtlcoder}, which focus on fine-tuning LLMs using specialized datasets, have demonstrated remarkable results. These developments underscore the evolving landscape of LLM applications in HDL debugging, highlighting both achievements and areas ripe for further exploration.

\section{Conclusion}\label{sec:conclusion}
In this paper, we propose an LLM-assisted HDL debugging
framework, namely HDLdebugger, which consists of HDL debugging data generation,
a search engine, and a retrieval-augmented LLM fine-tuning approach. Through extensive and varied experimentation with multiple LLMs, we have unearthed pivotal findings within the domain. Our method significantly surpasses existing techniques, achieving an exceptional pass-rate of up to 81.93\%, indicating that HDLdebugger could automate and streamline HDL debugging for chip design.
In addition, we provide in-depth experimental analysis and outline potential future directions for HDL debugging with LLMs in Appendix~\ref{sub:analysis}.

\clearpage
\balance
\bibliographystyle{ACM-Reference-Format}
\bibliography{DFT}
\clearpage
\appendix

\section{Additional Materials for Data Generation}
\Cref{table:DRC_RULE} indicates the common error type in HDL code summarized by LLM. By locating the main cause of errors, we can design multiple modification functions to recur the errors in a given HDL code. It is easy to reveal that some simple and similar modifications can result in totally different errors when matching the error causes and modification functions. These modification functions then become an essential part of the reverse engineering pipeline and the foundation of solutions in the RAG search engine.

The error message and solution database of the RAG search engine is built on the foundation of the modification functions. \Cref{table:RAG-DRC_RULE} indicates information in the database. For each common error code in HDL code compilation, it provides related descriptions and error root reasons that help debug. And most essentially, the database collects recommending solutions for a given error code. When retrieving knowledge with the RAG search engine, the root reason and solution together with the error message and solution will be returned to form LLM input.

\begin{table*}[th!]
    \caption{Sample HDL error and modification functions.}
    \begin{threeparttable}
    \begin{tabular}{c| c|c}
    \toprule
                     Error Type & Error Cause & Modification Function  \\ \midrule
    Memory        &  RAM connection or RAM port             &  Pulse signal modification                             \\
    Clock       &    Clock unit connection and availability          &  Pulse signal or assignment modification                              \\
    Trace        &    Trace unit availability and usage                 &  Assignment modification                                   \\ 
    Compression       &  Compression unit definition and I/O              &   Register or assignment modification                                  \\ 
    Data       &  Data Value                   &  Register or pulse signal modification                                   \\
    Test Point       &   Procedures completeness                  & All variables modification                                    \\
    Pattern Reform       & Pattern equivalence                     &  Assignment and probe modification                                    \\
    Syntax E       &    Sscript syntax error                 &   Syntax modification                                  \\
    Netlist E       &    Netlist attribute                 &   Netlist modification                                  \\
    Simulation E   &  Simulation config                   &   Config modification                                \\ \bottomrule
    \end{tabular}
    \begin{tablenotes}
    \footnotesize
    \item[1] The errors here are abstracted and modified for information security, but still reflect the categories and causes without revealing details.
    \end{tablenotes}
    \end{threeparttable}
    \label{table:DRC_RULE}
\end{table*}

\begin{table*}[th!]
    \caption{Examples of error information.}
    \begin{threeparttable}
    \begin{tabular}{c| c|c|c}
    \toprule
                     Error code & Descriptions & Root-Reasons & Solutions \\ \midrule
    C-error-1        &  Netlist not correctly obtain defined signal            &  Definition not match netlist & Modify definition or netlist                            \\
    C-error-2       &    Not define top port as output          &  Signal not output & Modify top port as output                             \\
    C-error-6        &    Clock signal not off during **                 &  ** get wrong signal  & Correctly off clock signal                                \\ 
    T-error-2       &  Clock definition duplicate              &   clock name conflict & Delete one definition                                  \\ 
    T-error-4       &  Probe initilized as 0 in **                   &  Invalid initialization     & Modify probe initialization to non-0                             \\
    T-error-18       &   Assignment in non-initialization stage      & Wrong assignment      &  Delete assignment                          \\
    T-error-27       & Pulse non-exist variable  &  wrong pulse & Delete  pulse                                \\
    M-error-1       &    Memory close when read                 &  Wrongly off memory  &  Correct memory and constraint              \\
    M-error-17       &    Logical loop in netlist                 &   Logical loop in netlist        &  Modify netlist              \\
    P-error-8   &  Definition lack of shift                   &   Wrong definition      &      Modify definition            \\ \bottomrule
    \end{tabular}
    \begin{tablenotes}
    \footnotesize
    \item[1] The rules here are abstracted and modified for information security, but still reflect the meanings, causes and solutions without revealing details.
    \end{tablenotes}
    \end{threeparttable}
    \label{table:RAG-DRC_RULE}
\end{table*}

\section{Additional Materials for RAG}
\subsection{BERT-LSTM}\label{appx:bert_lstm}
The BERT-LSTM model synergizes the contextual embedding capabilities of BERT (Bidirectional Encoder Representations from Transformers) with the sequential data processing strength of LSTM (Long Short-Term Memory) networks.
Specifically, given a buggy code $b$ and its error message $m$, we first concatenate them into a single sequence $qs=[\mathrm{CLS}, b, \mathrm{SEP},m]$. 
Since the token length of buggy code and error message is too long, we separate the query sequence $qs$ into a set of sub-sequences $\{s_k\}_{k=1}^{|qs|/n_s}$, where each subsequence has $n_s$ tokens.
Then, we fed each subsequence $s_k$ into BERT, and BERT will output a sentence-level vector 
$\mathbf{h}_k=BERT(s_k)$, where $\mathbf{h}_k$ 
captures the context information of tokens in $s_k$.
After obtaining all subsequence embeddings $\{\mathbf{h}_k\}_{k=1}^{|qs|/n_s}$ for $\{s_k\}_{k=1}^{|qs|/n_s}$, 
we employ a Bi-LSTM to capture long-range dependencies and intricate patterns  across different subsequences as $\{\mathbf{h}^{'}_k\}_{k=1}^{|qs|/n_s}=\mathrm{BiLSTM}(\{\mathbf{h}_k\}_{k=1}^{|qs|/n_s})$.
Upon acquiring the sequence of Bi-LSTM outputs, we use a self-attention mechanism to generate the final representation of $qs$ as $\mathbf{z}^s=\mathrm{Self\_Attention}(\{\mathbf{h}^{'}_k\}_{k=1}^{|q|/n_s})$.
The self-attention blocks attend to different parts of the LSTM sequence, enabling the model to focus on the most relevant features for classification.
Then, we feed $\mathbf{z}^s$ to a linear layer to predict buggy pattern probability for the buggy code query $(b,e)$.

\subsection{Greedy Algorithm}\label{appx:greedy_algorithm}
As introduced in Sec.~\ref{ssec:search_engine},  the top-$k$ buggy code problem in Eq.~\eqref{eq:rerank_obj} is \textit{NP-hard},  indicating that we cannot obtain the optimal top-$k$ buggy code instances $D^q_c$
in any polynomial time.
Thus, we propose a greedy algorithm in  Alg.~\ref{alg:ranker} with a theoretical guarantee to optimize the re-rank objective in Eq.~\eqref{eq:rerank_obj}.
For clarification, we denote the Eq.~\eqref{eq:rerank_obj} by $S(D^q_c)$ as follows.
$$S(D^q_c)=\max_{D^q_c \subseteq \hat{D}^q_c}{\sum_{I_i \in D^q_c}{sim(q,I_i)} +  \frac{1}{k}\cdot \sum_{I_i \in D^q_c}{dis(I_i, D^q_c)} }$$ 
The basic idea of Alg.~\ref{alg:ranker} is to greedily select the buggy code instance that can bring the maximum information gain into the selected instance set $D^q_c$ until it exceeds the number $k$.
Specifically, given the instance set $D^q_c$, we first define the marginal information gain of $I_j$ as follows:
\begin{equation}\label{eq:mar_score}
 \triangle S(I_j|D^q_c) = S(I_j\cup D^q_c) - S(D^q_c)
\end{equation}

As illustrated in Alg.~\ref{alg:ranker}, we first initialize the instance set $D^q_c$ as $\emptyset$ (line 1).
Then, we compute the similarity score $sim(q,I_j)$ for each $I_j \in \hat{D}^q_c$ (lines 2-4). 
Next, we compute the marginal information gain of each $I_j \in \hat{D}^q_c$ and 
select node $I^*$ with the maximum $ \triangle S(I_j|D^q_c)$ following Eq.~\eqref{eq:mar_score} (line 6-9).
Then, we
add $I^*$ into ${D}^q_c$, and remove it from $\hat{D}^q_c$ (lines 10-11). 
We repeat the selection procedure until we have selected $k$ buggy code instances (lines 5-12).

\begin{theorem}\label{theo:reranking}
Alg.~\ref{alg:ranker} can achieve a $1-1/e$ approximation ratio.
\end{theorem}

\begin{proof}
	The  $S(D^q_c)$ in Eq.~\eqref{eq:rerank_obj} is monotone and submodular.
 \begin{itemize}[leftmargin=12pt]
     \item {Monotone: } 
     Given any $I_i \in \hat{D}^c_q$ and selected instances set $D^c_q$, we can obtain $S(D^q_c\cup I_i)-S(D^q_c)\ge 0$.
     Thus, $S(D^q_c)$ is 
     monotone increasing.
     \item {Submodularity: }
     Given  ${\tilde{D}}^q_c \subset {D}^q_c$, and a buggy code instance $I_i \notin \tilde{D}^q_c, {D}^q_c$, 
    we can obtain:
    \begin{align}
        S(\tilde{D}^q_c \cup I_i) -  S(\tilde{D}^q_c) = sim(q,I_i) + \frac{1}{k} dis(I_i, \tilde{D}^q_c) \\
        S({D}^q_c \cup I_i) -  S({D}^q_c) = sim(q,I_i) + \frac{1}{k} dis(I_i, {D}^q_c) 
    \end{align}
    Since $dis(I_i, \tilde{D}^q_c)=\min_{I_j \in \tilde{D}^q_c }(2-sim(I_i,I_j))$ and ${\tilde{D}}^q_c \subset {D}^q_c$,
    thus we can obtain the inequality as:
    $$
     S(\tilde{D}^q_c \cup I_i) -  S(\tilde{D}^q_c) \ge  S({D}^q_c \cup I_i) -  S({D}^q_c).
    $$
    It demonstrates that $S({D}^q_c)$ in Eq.~\eqref{eq:rerank_obj} is submodular.
 \end{itemize}
	
 Since $S({D}^q_c)$ is monotone and submodular, according to~\cite{hochbaum1996approximating}, the  approximation ratio of Alg.~\ref{alg:ranker} is $1-1/e$.
\end{proof}

\fakeparagraph{Time Complexity}
Assume the dimension of $\mathbf{z}^w$ and $\mathbf{z}^s$ are $d_w$ and $d_s$, respectively.
First, it takes $O(N(d_w+d_s))$ to compute the similarity score $sim(q,I_j)$ for each $I_j \in \hat{D}^q_c$ (line 2-4). 
Then, it takes $O(k^2(d_w+d_s))$ to select the top-$k$ buggy code instances from $\hat{D}^q_c$ (line 5-12). 
Thus, the time complexity of the top-$k$ buggy code selection algorithm is  $O((N+k^2)(d_w+d_s))$ in total.

\section{Retrieval-augmented LLM Fine-tuning}
 
\subsection{Implementation Details }\label{ssec:impl} 
We implement our approach in PyTorch~\cite{paszke2019pytorch}, and fine-tune on CodeLlama~\cite{roziere2023code} which is provided by huggingface~\cite{wolf2019huggingface} model zoo. 
The foundation of our code is built upon the FastChat and Alpaca frameworks, incorporating cutting-edge technologies such as flashattention~\cite{dao2022flashattention}, deepspeed~\cite{rasley2020deepspeed}, to enhance effectiveness during both training and inference phases. 
Our experimental setup utilizes eight NVIDIA-GTX A100 GPUs with 80G memory to ensure enough computational capacity.
For training, we primarily adhere to the default hyperparameters. 
During the inference stage, we employ a greedy decoding strategy, akin to the approach used in ChipNemo~\cite{liu2023chipnemo}, to mitigate the significant compilation costs associated with this process.

\begin{algorithm}[t]
    \caption{Top-$k$ relevant code selection}
    \begin{algorithmic}[1]
    \Require  Buggy code query $q=(b,e)$ and $N$ code instances $\hat{D}^q_c$, and parameter $k$.
    \Ensure Top-$k$ instances $D^q_c \subseteq \hat{D}^q_c$.

    \State \textbf{Initialize:} $\hat{D}^q_c \gets \emptyset.$
    
    \For{$I_j \in \hat{D}^q_c$}
        \State $Sim(q,I_i) \gets \mathbf{Eq.}~\eqref{eq:sim_score}$  
    \EndFor
        
    \For{$i = 1$ \textbf{to} $k$}
        \For{$I_j \in \hat{D}^q_c$}
            \State $\triangle S(I_j|D^q_c) = S(I_j\cup D^q_c) - S(D^q_c)$  
        \EndFor
        \State $I^* = \arg\max_{I_j \in \hat{D}^q_c}{\triangle S(I_j|D^q_c)}$
        \State $\hat{D}^q_c = \hat{D}^q_c \setminus I^*$
        \State ${D}^q_c = {D}^q_c \cup I^*$
    \EndFor
    
    \State \Return ${D}^q_c$.
    
    \end{algorithmic}
    \label{alg:ranker}
\end{algorithm}

\begin{algorithm}
    \caption{Thoughts Generation Flow}
    \begin{algorithmic}[1]
    \Require Thoughts generation prompt $p_t$, correct response prompt $p_c$, buggy code $b$, correct code $c$, error message as $m$, retrieved information as $rag^d$, large language model as $LLM$, Number of thoughts to generate $k$.
    \Ensure Generated thoughts $\mathbf{t_i}$.
    
    \State \textbf{Initialize:} $G \gets \{\}$, $C \gets \{\}$, $D \gets \{\}$.
    
    \For{$j = 1$ \textbf{to} $k$}
        \State $t_j \gets LLM(p_t \circ b \circ c  \circ m \circ rag^d)$ \texttt{//Generate thoughts}. 
        \State $G \gets G \cup t_j$ \texttt{//Append $t_j$ to $G$}. 
    \EndFor
    
    \For{\texttt{each} $t_j$ \texttt{in} $G$}
        \State $c_j \gets LLM(p_c \circ b \circ m \circ rag^d \circ t_j)$ \texttt{//Predicted correct script}. 
        \State $C \gets C \cup c_j$ \texttt{//Append $c_i$ to $C$}. 
    \EndFor
    
    \For{\texttt{each} $c_j$ \texttt{in} $C$}
        \State Calculate edit distance $d_j \gets \text{EditDistance}(c, c_j)$
            \State $D \gets D \cup d_j$ \texttt{//Append $d_j$ to $D$}. 
    \EndFor
    
    \State Sort $D$ and obtain related thoughts $t^*_j$,
    \State \Return $t^*_j$.
    
    \end{algorithmic}
    \label{alg:method:thoughts-generation}
\end{algorithm}

\begin{figure}[th!]
    \centering
    \includegraphics[width=0.9\linewidth]{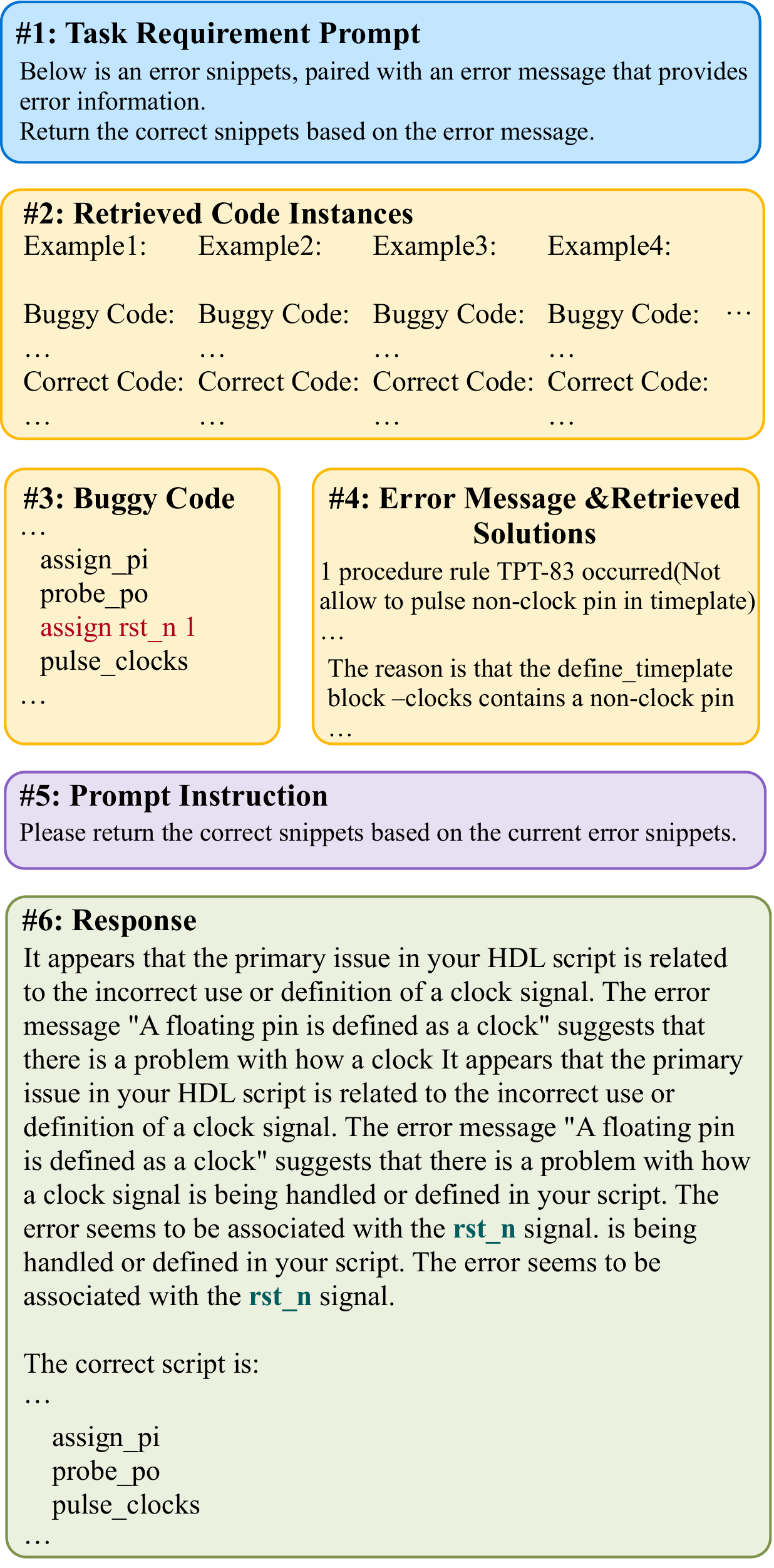}
    \caption{A case of HDL debugging.}
    \label{fig:example-hdl-debugging}
\end{figure}

\subsection{Thoughts Generation}
As \Cref{alg:method:thoughts-generation} illustrate the flow of thoughts generation, 
For simplicity, we omit the sample index.
The process mainly consist of 4 major steps. 
We firstly concatenate Thoughts generation prompt $p_t$, Buggy code $b$, Correct code $c$, Error message $m$, relevant information $rag^d$, and pass to LLM for inference, result of which stands for generated thoughts $t_1\dots t_k$. After obtaining the thoughts for the error code, we then concatenate correct response prompt $p_c$, Buggy code $b$, Error message $m$, , and generated thought $t_j$ to construct the step 2 input for LLM. 
The inference result denotes the predicted correct code script $c_1\dots c_k$ for correction the given error code. Given above two steps, the code correction is prepared successfully. The rest steps will focus on finding the best suited correction for given error code. To accomplish that, for given predicted correct code script $c_1\dots c_k$, we obtain the edit distance $d_1\dots d_k$ from each to original correct code $c$. The final procedure is to sort the edit distances and retrieve the generated thought $t^*_j$ corresponding to the lowest edit distance $d_j$. The final $t^*_j$ serves as our though output for the flow.

We delve deeper into the process of self-guided thought generation
The core principle of our approach involves submitting both the buggy code and its corrected version to the LLM, subsequently prompting the model to generate instructive guidance. 
We denote the correct code as $c_i$ and thoughts as $t_i$. 
Generally, compute $p(c_i|t_i)$ proves challenging due to the inaccessibility of suitable thoughts.
By leveraging Bayes' rule, we can reparameterize the formulation  as $p(c_i|t_i) \propto p(t_i|c_i)p(c_i)$. 
By focusing on $p(t_i|c_i)$, we encourage the LLM to align its outputs with verified correct code.

\subsection{Temperature setting}
By modifying temperature we can generate multiple thoughts differently.
We use multinomial sampling to generate samples randomly.
Define a sequence of input tokens $\mathbf{x}=\{x_1, x_2, ..., x_n\}$, denote $\pi_{\theta}$ as the LLM generation without passing final softmax layer. The next token $x_{n+1}$ can be obtained by:
\begin{equation}
    \begin{aligned}
    out ={}& \pi_{\theta}(\mathbf{x}) \\
    pr ={}& softmax_{T}(out) \\
    x_{n+1} \sim{}& Categorical(pr),
    \end{aligned}
\end{equation}
where $out$ is the next token logits output. 
$softmax_T$ means the softmax function with temperature where the formula of probability $pr_i$ is $\frac{e^{out_i} / T}{\sum_{j}e^{out_j} / T}$.
$T$ is the temperature parameter where higher $T$ makes the output distribution more uniform, thus introducing more randomness.
$Categorical$ means Categorical distribution.
For example, assume the next token probability distribution is $\{dog:0.4,cat:0.5,bike:0.1\}$, then the next token is selected according to their probabilities.
We can generate $K$ thoughts by this sampling strategy.
Therefore, we can generate thoughts randomly by controlling the temperature parameter.

\subsection{Debugging example}
\Cref{fig:example-hdl-debugging} illustrates an example of HDL debugging.
When given an error script and related error messages, we first use our search engine to retrieved related expert solutions and similar in-context samples (default 5 buggy code instances).
Then we feed buggy codes, error messages, and related in-context samples and expert solutions to our fine-tuned LLMs.
Finally, the LLM will produce a corresponding analysis and related correct script.

\section{Analysis and Discussion} \label{sub:analysis}
This section delves into a detailed examination of our research findings, focusing on the utility and implications of leveraging LLMs for debugging within HDL environments. We provide a thorough assessment of LLM-based debugging capabilities and the effects of iterative debugging procedures. Additionally, we explore both the challenges and prospects of incorporating LLMs into the HDL debugging framework.

\subsection{Trade-offs in RAG and SFT}Typically, in a Retrieval-Augmented Generation (RAG) system, the LLM acts as an agent that remains untuned to preserve its generalization ability, which might be compromised by task-specific parameter optimization. 
However, to enhance its efficacy in debugging, we find it essential to fine-tune the LLM, which presents a significant challenge as LLMs are expected to address a broader array of problems. 
One solution  incorporate a wider range of common instructional data during training to retain its general applicability. 
Another innovative strategy involves constructing a multi-LLM architecture housing numerous "expert" models to tackle specific domain challenges. 
Additionally, HDL debugging encompasses various tasks across different stages of electronic design automation, prompting us to consider developing a comprehensive framework in future investigations.

\subsection{Iterative Debugging}Our investigation predominantly concentrates on single-round debugging, noting that existing studies, such as~\cite{tsai2023rtlfixer}, demonstrate that a vast majority of issues (about 90\%) are resolvable in a single iteration. Nonetheless, complex scenarios necessitate multiple debugging rounds, posing a substantial challenge due to the high costs associated with gathering multi-round data. 
Moreover, effective iterative debugging demands LLMs capable of enhanced contextual analysis and comprehension. 
This area will be a focal point of our subsequent research endeavors.

\subsection{Enhancing the User Experience in Debugging Systems}
Within our debugging framework, we can quantitatively assess performance metrics such as pass rates and execution times. 
However, during the deployment phase, we observed that traditional debugging approaches—transforming erroneous scripts into correct ones are insufficient. 
In the context of HDL, the emphasis extends beyond mere error correction to enhancing the quality of the implementation, given the direct impact of hardware language on chip performance. 
Code that successfully compiles may not necessarily optimize chip design performance. 
Additionally, exact solutions are not always required; engineers often benefit from suggestive "hints" that inspire solutions to complex issues. 
Nonetheless, developing a metric to evaluate the quality of these hints represents a significant challenge.


\end{document}